\documentclass[prd,twocolumn,amsmath,amsfonts,amssymb,showpacs]{revtex4}
\usepackage{epsfig,psfrag,graphicx,bm,color,verbatim}

\def\msun{{\rm\,M_\odot}}

\def\sr{{\rm\,sr}}

\def\cm{{\rm\,cm}}

\def\kpc{{\rm\,kpc}}
\def\pc{{\rm\,pc}}

\def\GeV{{\rm\,GeV}}

\def\rsun{{R_\odot}}

\begin{document}
\title{Anisotropy probe of galactic and extra-galatic Dark Matter annihilations} 
\author{Mattia Fornasa}
\email{mfornasa@pd.infn.it}
\affiliation{University of Padova \& INFN sezione di Padova, via Marzolo 8, 35131 Padova, Italy}
\affiliation{Institut d'Astrophysique de Paris, boulevard Arago 89bis, 75014 Paris, France}
\author{Lidia Pieri}
\email{pieri@iap.fr}
\affiliation{Institut d'Astrophysique de Paris, UMR 7095-CNRS, Universit\'e Pierre et Marie Curie, boulevard Arago 89bis, 75014 Paris, France} 
\author{Gianfranco Bertone}
\email{bertone@iap.fr}
\affiliation{Institut d'Astrophysique de Paris, UMR 7095-CNRS, Universit\'e Pierre et Marie Curie, boulevard Arago 89bis, 75014 Paris, France} 
\author{Enzo Branchini}
\email{branchin@fis.uniroma3.it}
\affiliation{Department of Physics, Universit\`a di Roma Tre, via della Vasca Navale 84, 00146, Rome, Italy}

\begin{abstract}
We study the flux and the angular power spectrum of gamma-rays 
produced by Dark Matter annihilations in the Milky Way (MW) and in 
extra-galactic halos. The annihilation signal receives contributions from: 
a) the smooth MW halo, b) resolved and unresolved substructures in the MW, 
c) external DM halos at all redshifts, including d) their substructures. 
Adopting a self-consistent description of local and extra-galactic 
substructures, we show that the annihilation flux from substructures in the 
MW dominates over all the other components for angles larger than $\cal{O}$(1) 
degrees from the Galactic Center, unless an extreme prescription  is adopted 
for the substructures concentration. We also compute the angular power 
spectrum of gamma-ray anisotropies and find that, for an optimistic choice of 
the particle physics parameters, an interesting signature of DM annihilations
could soon be discovered by the {\small Fermi LAT} satellite at low 
multipoles, $\ell \lesssim 100$, where the dominant contribution comes from 
MW substructures with mass $M \gtrsim 10^4 M_{\odot}$. For the substructures 
models we have adopted, we find that the contribution of extra-galactic 
annihilations is instead negligible at all scales.
\end{abstract}

\maketitle

\section{Introduction}
\label{sec:one}
Despite the compelling evidence for Dark Matter (DM), we know very little 
about its nature. It is commonly assumed that DM is composed of Weakly 
Interacting Massive Particles (WIMPs), that are kept in thermal and kinetic 
equilibrium with baryons in the early universe through their weak coupling with
ordinary matter, and that subsequently decouple from them when the 
self-annihilation rate drops below the expansion rate of the universe, thus 
naturally achieving the appropriate relic density. Being proportional to 
the square of the DM number density, the self-annihilation rate is today 
very small, but it can still lead to detectable signals in
regions with very high DM density~\cite{Bertone:2004pz,Bergstrom:2000pn}.

Indirect DM searches are based on the detection of particles originating 
from DM annihilation or decay. If one focuses on the most widely 
discussed candidates, i.e. the supersymmetric neutralino and the 
so-called $B^{(1)}$ in theories with Universal Extra Dimensions, the mass of 
the DM particles should be approximately in the 100 GeV -- 1 TeV range, and 
the characteristic energy of the produced particles, roughly an order of 
magnitude smaller. At these energies, photons are among the best messengers, 
since gamma-rays travel in the local universe along geodesics without
significant energy losses.

The most obvious target for indirect DM searches is the center of our Galaxy,
where the large concentration of ordinary matter is believed to
be associated to a large DM density enhancement. However, the search 
for gamma-rays from DM annihilations in the Galactic Center (GC) is 
complicated by the presence of a strong signal of astrophysical origin, 
due to the point sources detected by the {\small EGRET} satellite 
\cite{MayerHasselwander:1998hg,Cesarini:2003nr,Hooper:2002ru} and by the 
{\small H.E.S.S.} telescope \cite{Aharonian:2006wh}. A further degree of 
complication is constituted by theoretical uncertainties on the DM profile in 
the innermost regions of the MW halo which is inevitably affected by the 
presence of a SuperMassive Black Hole and by the surrounding distribution of 
stars ~\cite{Bertone:2005hw,Gondolo:1999ef}.

To avoid such difficulties, different authors have suggested alternative
strategies to unambiguously detect the signature of DM interpretation.
One possibility is to look for enhancements in the gamma-ray flux with
no obvious astrophysical origin. Moreover, the presence of possible 
distinctive features in the gamma-ray energy spectrum (as lines 
\cite{Bergstrom:1997fj,Bergstrom:2004nr} or ``bumps'' \cite{Bringmann:2007nk}) 
may help in ruling out alternative, more conventional, interpretations.
Ando et al. \cite{Ando:2005xg,Ando:2006cr} have recently pointed out that 
unique annihilation features may also be detected in the angular 
correlation properties of the Extragalactic Gamma-ray Background (EGB).
From an observational point of view the EGB is obtained by subtracting
the galactic gamma-ray flux produced by cosmic-rays interacting with the 
interstellar medium at high galactic latitude from the total background 
measured by {\small EGRET} \cite{Strong:1998fr}. The resulting EGB is 
isotropic and its energy spectrum is rather uncertain, especially at energies 
above 10 GeV \cite{Sreekumar:1997un,Strong:2004de,Stecker:2007xp}. Unresolved 
extra-galactic gamma-ray sources like blazars certainly contribute to the EGB. 
However, current uncertainties on the luminosity function of these objects do 
not allow to exclude additional contributions from more exotic processes like 
DM annihilation within and outside our Galaxy.

The annihilation signal along a given direction in the sky is contributed 
by four different sources:
\begin{itemize}
\item the smooth MW halo, that is expected to contribute most towards
the GC where the DM density enhancement is expected to be very large (see 
references above),
\item resolved and unresolved substructures within the MW halo (see 
\cite{Pieri:2007ir} and references therein).
\item extra-galatic DM halos at all redshifts \cite{Ullio:2002pj}, including
\item their substructures.
\end{itemize}

Ando et al. have shown that if DM annihilations in extra-galactic
halos and their substructures contribute more than 30\% to the EGB, the 
recently launched {\small Fermi LAT} satellite should be able to identify 
this 'exotic' contribution in the angular power spectrum.

Their analysis does not account for the possible contribution of galactic 
substructures to the angular correlation signal 
\cite{Ando:2005xg,Ando:2006cr}. Such contribution has been studied recently 
by Siegal-Gaskins who concluded that detectable features in the EGB angular 
power spectrum may exist and they can be used to infer the presence of 
galactic substuctures and constrain their abundance \cite{SiegalGaskins:2008ge}
(in this case, annihilations in extra-galactic halos have been neglected).

Our model for galactic substructures is taken from Ref. \cite{Pieri:2007ir}, 
which is inspired by the high resolution N-body simulations aimed, precisely, 
at resolving subclumps within galaxy-sized halos (see 
Refs. \cite{Diemand:2005vz,Diemand:2006ik} and the more recent Refs.
\cite{Diemand:2008in,Springel:2008by,Springel:2008cc}). At $ z=0 $, 
substructures are predicted to span 12 orders of magnitude in mass. Thus, the
description of smaller objects is based on the extrapolation from the
results at higher masses, since resolving such small clumps would be well 
beyond the resolution limit of the N-body experiments.
In the attempt of limiting the theoretical uncertainties involved in this 
extrapolation, we have adopted two different prescriptions for the 
concentration of DM halos and subhalos of mass $M$ as a function of redshift 
$z$, $c(M,z)$.
Then, to compute the gamma-ray fluxes from the resulting $\sim 10^{16}$ 
galactic substructures, we have used the hybrid analytical and Monte Carlo 
approach described in Ref. \cite{Pieri:2007ir}. 

The extra-galactic flux has been modeled using the formalism of Ref. 
\cite{Ullio:2002pj}, adapted to our calculation setup. In this way, we were 
able to compute for the first time in a self-consistent way the normalization 
of the DM extra-galactic signal relative to the DM galactic foreground, a 
quantity that does {\it not} depend explicitly on the particle physics 
parameters (a weak implicit dependence remains, since the lower limit of 
integration of the subhalo mass function is sensitive to the particular
particle physics scenario).

We have then computed the angular power spectra by running the public code 
{\bf HEALPix} on our mock maps of galactic substructures, and comparing our 
result to those of Ref. \cite{Ando:2006cr} and Ref. \cite{SiegalGaskins:2008ge}.

The paper is organized as follows: in Sections \ref{sec:two}-\ref{sec:four} 
we describe the theoretical setup, the properties of our model substructures 
and compute the mock full-sky maps of the gamma-ray flux due to DM 
annihilations, both galactic and extra-galactic. In Section \ref{sec:five} we 
compute the angular power spectrum of the mock gamma-ray sky and simulate
its measurement using {\small Fermi LAT} in Section \ref{sec:six}.
Finally, in  Section \ref{sec:seven} we draw our main conclusions and
discuss our results.

\section{Simulation of galactic substructures}
\label{sec:two}

The Cold Dark Matter scenario predicts the formation of a large number 
of DM halos, virialized structures with masses possibly as small as 
$\sim 10^{-6} \msun$ \cite{Green:2003un,Green:2005fa}.
Such halos merge into larger and larger systems, leading to the formation of 
today's halos of galaxies and clusters of galaxies. A fraction of the 
small halos survives to dynamical interactions with the stellar and dark 
components until the present epoch.
Earth-size substructures have indeed been found in numerical simulations 
within DM halos of mass $\sim 0.1 \msun$ at a redshift of 74 
\cite{Diemand:2005vz}.

Resolving such small structures within a galaxy-sized halo today is out of 
computational reach, so that their spatial distribution, mass function and 
internal structure can only be estimated upon a rather uncertain extrapolation 
from the properties of higher mass satellites.
Here, we work under the following hypotheses:
\begin{itemize}
\item  we assume that substructures trace the mass of the MW, i.e that their 
radial distribution follows the mass density profile of the parent halo (as 
found e.g. in Ref. \cite{Diemand:2005vz}, see however the discussion below on 
the results of the most recent numerical simulations),
\item we assume that the substructures mass function is well approximated by 
a power-law ${\rm d} n(M)/{\rm d ln}(M) \propto M^{-1}$, normalized so that 
10\% of the MW mass lies in objects within the mass range 
$ 10^{-5} M_{\rm MW}-10^{-2} M_{\rm MW}$, as found in Ref. \cite{Diemand:2006ey}.
\end{itemize}

Under these assumptions, the number density of subhalos per unit mass at a 
distance $R$ from the GC can be written as:   
\begin{equation}  
\rho_{sh}(M,R) = A M^{-2} \frac{\theta (R - r_{min}(M))}
{(R/r_s^{\rm MW}) (1 + R/r_s^{\rm MW})^{2}} \msun^{-1}  \kpc^{-3}, 
\label{rho}  
\end{equation}  
where $r_s^{\rm MW}$ is the scale radius of our Galaxy and the effect of tidal 
disruption is accounted for by the Heaviside step function 
$\theta (R - r_{min}(M))$.
The tidal radius, $r_{min}(M)$, is found by using the Roche criterion to decide 
whether a subhalo of mass $M$ survives tidal interaction with the host halo
(see Ref. \cite{Pieri:2007ir}). 
According to our normalization about 53\% of the MW mass is condensed within 
$\sim 1.5 \times 10^{16}$ subhalos with masses in the range 
$10^{-6}-10^{10} \msun$. Their abundance in the solar neighborhood turns out 
to be $\sim 100 \mbox{ subhalos }\pc^{-3}$. 

As far as the smooth component of the Galactic halo is concerned, we assume 
that the MW halo follows a Navarro, Frenk \& White (NFW) 
profile \cite{Navarro:1995iw,Navarro:1996gj}, with a scale radius 
$r_s^{\rm MW}=21.7 \kpc$ and that the total mass enclosed in a radius 
$r_{200}$, corresponding to the radius where the halo density is $200$ 
times the critical density of the universe, is $M_{\rm MW}=10^{12} \msun$.
An NFW fit to the DM halo of our Galaxy is consistent within 10\% with the 
results of the {\it Via Lactea I} simulation \cite{Diemand:2006ik}.
A popular way to characterize the mass distribution within the DM halo is  
using the so-called \emph{concentration}, a shape parameter defined as the 
ratio between the virial radius and the scale radius, 
$c \equiv r_{200}/r_s $.

Subhalos are also assumed to follow an NFW profile, with a concentration that 
depends on their mass \cite{Diemand:2005vz}. To determine the dependence of 
concentration on the mass, we follow the prescription proposed in 
Ref. \cite{Bullock:1999he} (hereafter B01) in which the concentration 
parameter is found to depend on both the subhalo mass $M_{h}$ and on its 
collapse redshift, $z_{\rm c}$, defined as the epoch in which a mass scale 
$M_{h}$ enters the non-linear regime.
The collapse occurs when $\sigma(M_{h})D(z_{\rm c}) \sim 1$, where
$\sigma(M_{h})$ is the present linear theory amplitude of mass fluctuations 
on the scale $M_{h}$ and $D(z_{\rm c})$ is the linear theory growth factor 
at the redshift $z_{\rm c}$.
In an attempt to bracket our theoretical uncertainties in modeling 
$c(M_{h},z_{\rm c})$ and its extrapolation at low masses we have implemented 
two rather extreme models chosen among those considered in Ref. 
\cite{Pieri:2007ir}, namely:
\begin{itemize}
\item $B_{z_0,ref}$: that extrapolates the $c(M_{h},z_{\rm c})$
relation of B01 below  $10^4 \msun$ with a simple power-law, 
\item $B_{z_f,ref}$: which assumes that surviving subhalos do not change their 
density profile since their formation. Thus the concentration parameter at 
$z_c$ can be obtained from the one at $z=0$ through 
$c(M_{h},z_c)=c(M_{h}, z=0) / (1+z_c)$.
The values of $c(M_{h},z=0)$ corresponds to those of the $B_{z_0,ref}$ case and
the collapse redshift $z_c$ is obtained by extrapolating the expression 
proposed by B01 below $10^4 \msun$.
\end{itemize}

Finally, the concentration parameters are not uniquely defined by the halo 
mass. Rather, they follow a log-normal distribution with dispersion 
$\sigma_c$ = 0.24 \cite{Bullock:1999he} and mean $ \bar{c}(M)$:
\begin{equation}
P(\bar{c}(M),c) = \frac{1}{\sqrt{2 \pi} \sigma_c c} \, 
e^{- \left ( \frac{\ln(c)-\ln(\bar{c}(M))} {\sqrt{2} \sigma_c} \right )^2}.
\end{equation}

\subsection{Recent highlights from numerical simulations}
Two new sets of very high resolution numerical experiments have been recently 
released, namely the {\it Via Lactea II} \cite{Diemand:2008in} and the 
{\it Aquarius} \cite{Springel:2008by,Springel:2008cc} simulations.
The main characteristics of the MW subhalo population in these two simulations 
are summarized in Table \ref{tab:smodels} and compared to those of our 
subhalos.
Among the main differences, we note that the 
mass fraction in substructures is a factor 2-6 smaller than 
that used in our model. Another difference is in the subhalo distribution 
which, in our case, trace the smooth mass distribution of the MW halo, 
$\rho_{\rm MW}(r)$, while Refs.~\cite{Springel:2008by,Springel:2008cc}
suggest to use an Einasto profile \cite{Einasto} and in 
{\it Via Lactea II} the innermost regions are best fitted by a power-law 
$(1+r)^{-2}$.

The concentration parameter of subhalos in the two simulations depends from 
the distance from the GC and, in the {\it Aquarius} case, follows the 
prescription in Ref. \cite{Neto:2007vq} (N07) rather than the B01 model. 
Finally, in {\it Aquarius}, the subhalo density profile is also parametrized
as an Einasto, rather than an NFW, profile.

These differences, in particular the fact that neither in 
{\it Via Lactea II} nor in {\it Aquarius} the spatial distribution of 
subhalos trace the mass of the parent halo, do have a significant impact on 
the angular power spectrum of the gamma-ray flux, as we will show in an 
upcoming publication. Here, we perform all calculations under the assumptions 
listed above, and discuss how the differences among the subhalo models in 
Table \ref{tab:smodels} are 
expected to affect the angular power spectrum (Section \ref{sec:seven}).

\begin{table}
\begin{center}
\begin{tabular}{c|c|c|c}
\hline 
& this paper & {\it Aquarius} & {\it Via Lactea II} \\ 
\hline
clumpiness $(< r_{200})$ &  53\% & 8\% & 26\% \\ 
\hline
$N_{subhalos}(< r_{200})$ & $1.5 \times 10^{16}$ & $2.3 \times 10^{14}$ & $7 \times 10^{15}$ \\
\hline
$dn/dM$  & $ \propto M^{-2}$ & $ \propto M^{-1.9}$ & $ \propto M^{-2}$ \\ 
\hline
$n_{sh}(R)$ & NFW & Einasto & $\propto (1+R)^{-2}$ \\  
& & $\alpha_{\rm Einasto}=0.68$ &  \\ 
\hline 
 $c(M,z)$ &  B01 &  N07 & B01 \\ 
\hline
$\rho_{\rm MW}(r)$ & NFW & Einasto & NFW \\
& & $\alpha_{\rm Einasto}=0.21$ & \\ 
\hline
$\rho_{halo}(R)$ & NFW & Einasto & NFW \\
& & $\alpha_{\rm Einasto}=0.16$ & \\ \hline
\end{tabular}
\caption{\label{tab:smodels} 
The main characteristics of the subhalo model considered in this work compared with those measured in the recent {\it Aquarius} and {\it Via Lactea II} N-body simulations. The clumpiness is defined as the fraction of the dark mass in substructures within the virial radius of the MW $r_{200} = 210 \kpc$. $N_{subhalos}$ is the total number of substructures within $r_{200}$. $dn/dM$ is the subhalo mass function, $n_{sh}(r)$ their radial distribution and $c(M,z)$ is the model used for the concentration. $\rho_{\rm MW}$ and $\rho_{halo}$ represent the mass density profile of the host halo and its substructures, respectively.}
\end{center}
\end{table}

\section{Modeling the gamma-ray flux}

\subsection{Gamma-ray flux from galactic subhalos}
\label{sec:three}
The gamma-ray flux expected from the annihilation of DM particles can be 
written as:
\begin{equation}
\frac{d \Phi_\gamma}{dE_\gamma}(E_\gamma,l,b) =
\frac{d \Phi^{\rm PP}} {dE_\gamma}(E_\gamma) \times \Phi^{\rm cosmo}(l,b)
\label{flussodef}
\end{equation}
where the term
\begin{equation}
\frac{d \Phi^{\rm PP}}{dE_\gamma}(E_\gamma) =  
  \frac{1}{4 \pi} \frac{\sigma_{\rm ann} v }{2 m^2_\chi} \cdot 
\sum_{f} \frac{d N^f_\gamma}{d E_\gamma} B_f  
\label{flussosusy}
\end{equation}
contains the dependence on particle physics parameters, while
$$ 
\Phi^{\rm cosmo}(\psi,l,b) = \int_M d M \int_c d c \int \int_{\Delta \Omega}
d \theta d \phi \int_{\rm l.o.s}  d\lambda
$$
$$
[ \rho_{sh}(M,R(\rsun, \lambda,l,b,\theta, \phi)) \times P(c) \times
$$
\begin{equation}
\times \Phi^{\rm cosmo}_{halo}(M,c,r(\lambda,l,b,\theta,\phi)) \times 
J(x,y,z|\lambda,\theta,\phi) ]
\label{smoothphicosmo}
\end{equation}
represents the contribution to the foreground emission from the subhalo 
population and
$$
\Phi^{\rm cosmo}_{halo}(M,c,r) = \int \int_{\Delta \Omega}  
d \phi ' d \theta '  \int_{\rm l.o.s} d\lambda '
$$
\begin{equation}
\left [ \frac{\rho_{halo}^2 (M,c,r(\lambda, \lambda ',l,b,\theta ' \phi '))} {\lambda^{2}} J(x,y,z|\lambda ',\theta ' \phi ') \right]
\label{singlehalophicosmo}
\end{equation}
is the contribution from a single subhalo.

The total galactic flux is obtained by adding to Eq. \ref{smoothphicosmo}
the contribution of DM annihilations in the smooth NFW halo of the MW.

In Eq. \ref{flussosusy}, $m_\chi$ is the DM particle mass, 
$\sigma_{\rm ann} v$ the annihilation cross section, and 
$d N^f_\gamma / dE_\gamma$ the differential photon spectrum per annihilation 
relative to the final state $f$, with branching ratio $B_f$ that we take 
from Ref. \cite{Fornengo:2004kj}.
In this work we adopt a rather optimistic particle physics scenario 
(in the sense that it provides large annihilation fluxes), in which 
the DM particle has a mass $ m_\chi = 40 \mbox{ GeV} $, a cross section
 $ \sigma v = 3 \times 10^{-26} \mbox{cm}^3 \mbox{s}^{-1} $ and 
particles annihilate entirely in $ b\bar{b}$.

In Eq. \ref{smoothphicosmo}, $\Delta \Omega=9.57 \times 10^{-6} \mbox{sr} $ 
is the solid angle considered for the integration, corresponding to the 
angular resolution of the {\small Fermi LAT} satellite, $ (l,b) $ are the 
galactic coordinates of the direction of observation, and $ J $ is the 
Jacobian determinant of the transformation between polar and cartesian 
coordinate systems. The galactocentric distance, $R$, can be written as a 
function of the coordinates inside the observation cone 
($\lambda$, $\theta$, $\phi$) and of $(l,b)$ through the relation 
$R = \sqrt{\lambda^2 + \rsun^2 -2 \lambda \rsun C}$, where $\rsun$ is the 
distance of the earth from the GC, and $C$ is the cosinus of the angle between
the direction of observation and the direction of the GC.

\subsection{Gamma-ray flux from extra-galactic structures}

Aside from local DM, the gamma-ray flux in any given direction receives a 
contribution from all the structures at all redshift along the line of sight. 
Adapting the formalism of Ref. \cite{Ullio:2002pj}, we estimate here the 
contribution of extra-galactic structures, including the presence of 
substructures following the $ B_{z_0,ref} $ and $B_{z_f,ref}$ models.

Following Ref. \cite{Ullio:2002pj}, the infinitesimal volume $ dV $ 
at a redshift $z$ can be written as
\begin{equation}
dV=\frac{R_0^3 r^2 dr d\Omega}{(1+z)^3},
\end{equation}
where $ d\Omega $ is the solid angle, $dr$ is the infinitesimal comoving depth
and $R_0$ is the scale factor at the present epoch.
If, for a moment, we neglect the presence of subhalos and assume that the 
gamma-ray emission is isotropic, we can compute the number 
$ d\mbox{\sffamily{N}}_\gamma $ of gamma-ray photons produced in $ dV $ in a 
time interval $ dt $ with an energy between $ E $ and $ E+dE $ and collected 
by a detector with effective area $ dA $ by integrating the single halo 
emissivity over the halo mass function $\frac{dn}{dM}(M,z)$: 
\begin{eqnarray}
\label{eqn:dN}
d\mbox{\sffamily{N}}_\gamma & = & e^{-\tau(z,E_0)} \left[ (1+z)^3 \int dM 
\frac{dn}{dM}(M,z) \right. \\ \nonumber 
& & \left. \frac{d\mathcal{N}_\gamma}{dE}(E,M,z) \frac{dV dA}{4\pi R_0^2r^2} 
dE_0 dt_0 \right],
\end{eqnarray}
where $ E_0 $ and $ dt_0 $ are, respectively, the energy and the time interval
over which the photons are detected on earth. These quantities are related 
to those at the redshift of emission through $ E_0=E/(1+z) $ and 
$ dt_0=(1+z)dt $, so that $ dt_0dE_0 = dtdE $. The halo mass function, 
$ dn/dM $, represents the comoving number density of DM halos of mass $ M $ 
at redshift $ z $ and the factor $ (1+z)^3 $ converts comoving into 
physical volumes. 
In the Press-Schechter formalism \cite{Press:1973iz}, the halo mass function 
is
\begin{equation}
\frac{dn}{dM}(M,z)=\frac{\rho_{\mbox{\tiny{cr}}}\Omega_{0,m}}{M^2}
\nu f(\nu) \frac{d\log\nu}{d\log M},
\label{eqn:dndM}
\end{equation}
where $ \rho_{\mbox{\tiny{cr}}} $ is the critical density, $\Omega_{0,m}$ is 
the mass density parameter, $\nu=\frac{\delta_{sc}(z)}{\sigma(M)}$, 
$ \sigma(M) $ is the rms density fluctuation on the mass scale $M$
and $\delta_{sc}$ represents the critical density for spherical collapse.
We refer to Refs. \cite{Ullio:2002pj}, \cite{Ahn:2007ty} and 
\cite{Eisenstein:1997jh} for the exact computation of these quantities.

$ d\mathcal{N}_\gamma/dE $ represents the number of photons with energy 
between $ E $ and $E+dE $ produced in a halo of mass $ M $ at redshift $ z $.  
The exponential $ e^{-\tau(z,E_0)} $ is an absorption coefficient that accounts 
for pair production due to the interaction of the gamma-ray photons with the 
extra-galactic background light in the optical and infrared bands. 
Following Ref. \cite{Bergstrom:2001jj}, we adopt here the following 
expression that accounts for current observational constraints:
$ \tau(z,E_0)=z/[3.3(E_0/10 \mbox{ GeV})^{-0.8}] $ . 
The mean flux is obtained by integrating Eq.~\ref{eqn:dN} along the 
line-of-sight:
\begin{eqnarray}
\label{eqn:dPhi_dE_extra}
& & \left\langle \frac{d\Phi}{dE_0d\Omega} \right\rangle(E_0) \equiv
\frac{d\mbox{\sffamily{N}}_\gamma}{dE_0 dt_0 dA d\Omega} \\ \nonumber
& = & \frac{1}{4\pi} \int dr R_0 e^{-\tau(z,E_0)} \\ \nonumber 
& & \int dM \frac{dn}{dM}(M,z) \frac{d\mathcal{N}_\gamma}{dE}(E_0(1+z),M,z) 
\\ \nonumber
& = & \frac{c}{4\pi} \int dz \frac{e^{-\tau(z,E_0)}}{H_0 h(z)} \\ \nonumber
& & \int dM \frac{dn}{dM}(M,z) \frac{d\mathcal{N}_\gamma}{dE}(E_0(1+z),M,z),
\end{eqnarray}
where the last expression has been obtained by transforming comoving 
distances $ r $ into redshifts $z$, through the introduction of the
Hubble parameter
$ H_0 h(z)= H_0 \sqrt{\Omega_{0,m}(1+z)^3+\Omega_{0,\Lambda}} $,
where  $ H_0 $ is the Hubble constant and $\Omega_{0,i}$ the abundance 
in units of the critical density at $z=0$. 

The number of photons emitted in a single halo, $ d\mathcal{N}_\gamma/dE $, 
depends on the DM density profile (NFW in our case) and on the particle 
physics scenario (particle mass $m_{\chi}$, annihilation cross section 
$\sigma v$ and differential energy spectrum per annihilation, 
$ dN_\gamma/dE_0 $). The NFW profile of a halo with mass $ M $ is completely 
specified by the concentration parameter and the virial overdensity 
$ \Delta_{vir} $ of the halo. The virial radius $ r_{vir} $, as opposed to 
the previously defined $ r_{200} $, is the radius of the sphere which 
enclosed an average density $ \Delta_{vir} \times \rho_m$.
To be consistent with the notation used in Ref. \cite{Ullio:2002pj} we define 
the concentration parameter as $ c(M,z)=r_{vir}/r_s $ and change the 
corresponding values in models $ B_{z_0,ref} $ and $ B_{z_f,ref} $ 
accordingly.

Bearing all this in mind, we can express $ d\mathcal{N}_\gamma/dE $ as
\begin{eqnarray}
\label{eqn:dmathcalNdE}
\frac{d\mathcal{N}_\gamma}{dE}(E,M,z) & = & \frac{\sigma v}{2} 
\frac{dN_\gamma(E)}{E} \frac{M}{m_\chi^2} 
\frac{\Delta_{vir} \rho_{\rm cr} \Omega_m(z)}{3} \\ \nonumber 
& & \frac{c^3(M,z)}{I_1(x_{\mbox{\tiny{min}}},c(M,z))^2} 
I_2(x_{\mbox{\tiny{min}}},c(M,z)). 
\end{eqnarray}
In the previous expression, the virial overdensity is \cite{Ullio:2002pj}:
\begin{equation}
\Delta_{vir}(z)=\frac{18 \pi^2 + 82(\Omega_m(z)-1) -39
(\Omega_m(z)-1)^2}{\Omega_m(z)}.
\end{equation}
and 
the integrals $ I_1 $ and $ I_2 $ have an analytic expression:
\begin{equation}
I_n(x_{\mbox{\tiny{min}}},x_{\mbox{\tiny{max}}})=\int g^n x^2 dx,
\end{equation}
where  $ g(x)=x^{-1}(1+x)^{-2} $.

In Eq. \ref{eqn:dmathcalNdE}, the lower integration limit is set at the 
minimum radius within which the annihilation rate equals the dynamical time:
$ x_{\mbox{\tiny{min}}}=10^{-8}\mbox{kpc}/r_s $ ($ r_s $ is the scale radius 
in kpc). We have checked that the results are not sensitive to a different 
choice for $ x_{\mbox{\tiny{min}}} $. 

Putting Eqs. \ref{eqn:dmathcalNdE} and \ref{eqn:dPhi_dE_extra} together, we 
obtain the expression for the isotropic gamma-ray flux from extra-galactic 
DM halos:
\begin{eqnarray}
\label{eqn:aveflux}
\left\langle \frac{d\Phi}{dE_0d\Omega}\right\rangle (E_0) = 
\frac{\sigma v}{8\pi}\frac{c}{H_0}
\frac{\rho_{\mbox{\tiny{cr}}}^2 \Omega_{0,m}^2}{m_\chi^2} \times  \ \ \ \ \ \ \ \ \ \  \\
 \int dz (1+z)^3 \frac{\Delta^2(z)}{h(z)} \frac{dN_\gamma(E_0(1+z))}{dE}  e^{-\tau(z,E_0)}, \nonumber
\end{eqnarray}
with
\begin{equation}
\Delta^2(z)=\int dM \frac{\nu(z,M)f(\nu(z,M))}{\sigma(M)} \left|
\frac{d\sigma}{dM} \right| \Delta_M^2(z,M)
\label{eqn:Delta2}
\end{equation}
and
\begin{equation}
\Delta_M^2(z,M)=\int dc^\prime P(c(M,z),c^\prime)
\frac{\Delta_{vir}}{3} \frac{I_2(x_{\mbox{\tiny{min}}},c^\prime)}
{I^2_1(x_{\mbox{\tiny{min}}},c^\prime)} (c^{\prime})^3 dc^\prime.
\label{eqn:Delta2M}
\end{equation}
$ \Delta_M^2(z,M) $ represents the enhancement in the gamma-ray flux due to 
the presence of a DM halo with a mass $ M $ at a redshift $ z $.
In $ \Delta^2(z) $ all these contributions are integrated over the halo mass 
function. Therefore $ \Delta^2(z) $ quantifies how much the annihilation
signal is boosted up by the existence of virialized DM halos.

To account for the presence of substructures in extra-galactic halos, 
we assume that a given fraction of the halo mass is concentrated in 
substructures with the same properties as the galactic subhalos described in 
Section \ref{sec:two}. For consistency with the MW case, we require  
10\% of the parent halo mass $ M $ to be in subhalos in the mass range  
$ 10^{-5}M - 10^{-2}M $. Following this requirement the mass fraction $f(M)$ in 
substructures within a host halo of mass $M$ can be fitted as 
\begin{equation}
f(M) \equiv \frac{M^{\mbox{\tiny{tot}}}_s}{M}=\frac{1}{6} 
+\frac{\log_{10}(M/M_\odot)}{30}.
\end{equation}

With the above definition, the presence of substructures is taken into account
replacing $ \Delta^2_M $ in Eq. \ref{eqn:Delta2M} with the following
expression:
\begin{eqnarray}
\label{eqn:substructures}
& \Delta_M^2(M) \rightarrow & (1-f(M))^2 \Delta_M^2(M)+ \\ 
& & + \frac{1}{M}\int dM_s M_s \frac{dn}{dM_s}(M_s,z) \Delta_{M_s}^2(z,M_s).
\nonumber
\end{eqnarray}

\section{Mapping the gamma-ray annihilation signal}
\label{sec:four}

\subsection{Galactic contributions}
\label{sec:fourpointone}
The gamma-ray flux from local substructures receives contributions from
all the subhalos along the line of sight, typically $\sim 10^9$ subhalos
when integrating over a a solid angle $ \sim 10^{-5} \mbox{sr} $. A 
brute-force integration with a Monte Carlo approach is therefore 
impossible even on high-speed computers. 
To circumvent this problem, Pieri et al. in Ref. \cite{Pieri:2007ir} have 
proposed a hybrid approach that consists in splitting the integral
into two different contributions. The first one, which we regard as due to 
unresolved subhalos, is the average contribution of a subhalo population 
distributed according to Eq. \ref{smoothphicosmo}, which can be estimated 
analytically. The second one comes from the nearest subhalos, that one may 
hope to resolve as individual structures, and it is estimated by numerical 
integration of 10 independent Monte Carlo realizations. 
To determine the number of individual subhalos in each Monte Carlo realization 
we assume that their contribution to the gamma-ray flux represents a Poisson 
fluctuation to the mean gamma-ray annihilation signal. 
Following this criterion, we only consider, in each mass decade, subhalos 
with $\Phi^{\rm cosmo} > \langle \Phi^{\rm cosmo}_{B_{z_0,ref}}
(\psi = 180^\circ) \rangle \sim  10^{-5} \GeV^2 \cm^{-6} \kpc \sr$, where the 
brackets indicate the mean annihilation flux.
For a given halo mass, this requirement corresponds to generate all subhalos 
within a maximum distance $ d_{max}(M) $. More than 500 subhalos are found 
within $ d_{max} $ for each mass decade from 10 $ M_\odot $ to $ 10^7 M_\odot $
and their positions are simulated according to Eq. \ref{smoothphicosmo}. 
For less massive subhalos, if less than 500 subhalos are found, we increase 
$ d_{max} $ to include all the nearest 500 subhalos in that mass range.

Pieri et al. in Ref. \cite{Pieri:2007ir} have checked that  
the predicted gamma-ray flux does not depend on $ d_{max} $. Here we 
need to perform a similar robustness test to check that the choice of 
$ d_{max} $ does not introduce spurious features in the angular power 
spectrum. For this purpose we have generated different Monte Carlo 
realizations simulating subhalos up to $ n \times d_{max} $ with $ n $ from 
2 to 5. We have checked that convergence in the power spectra is obtained 
already for $ n=2 $. Since the convergence depends weakly on the model 
concentration parameter adopted, in this paper we have adopted a more 
stringent criterion and set our maximum simulation distance at 
$ 3 \times d_{max}(M) $ to guarantee that the convergence is reached in 
both subhalo models considered.
 
To summarize, in our model we consider three sources of gamma-ray 
annihilation within the MW: the smooth Galactic halo (which we will refer to 
as NFW), the unresolved subhalos (UNRES), and the resolved substructures 
(RES). Since the total annihilation depends on the square of the DM density, 
for the purpose of computing the angular correlation signal we also need to 
consider the double products NFW $ \times $ RES and NFW $ \times $ UNRES. We 
neglect RES $\times $ UNRES since a resolved subhalo at a given location 
excludes the presence of unresolved substructures.

We point out that the subhalos labelled as "resolved" do not correspond to those
substructures that will be detected by {\small Fermi LAT}. Instead they
merely represent the nearest subhalos that we have generated to account for
the discreteness of the subhalo spatial distribution. Moreover, considering 
the population of RES subhalos also accounts for the pixel-to-pixel
variation in the annihilation flux that would be neglected focusing only on
the smooth UNRES signal.

The possibility of detecting galactic DM substructures is discussed in details 
in Ref. \cite{Pieri:2007ir}. Since, in model $ B_{z_0,ref} $, {\small Fermi LAT} 
is expected to detect only very few substructures, in this work all the 
simulated subhalos are included in the RES map. On the contrary, for model 
$ B_{z_f,ref} $, {\small Fermi LAT} may resolved up to $ \sim 100 $ point-like
sources. Since we are interested in the correlation properties of the
diffuse gamma-ray emission, these sources should be masked. However, we checked
that the angular spectrum for the model $ B_{z_f,ref} $ does not significantly
depend on the exclusion of these point-like sources.

\begin{figure}
\includegraphics[width=0.5\textwidth]{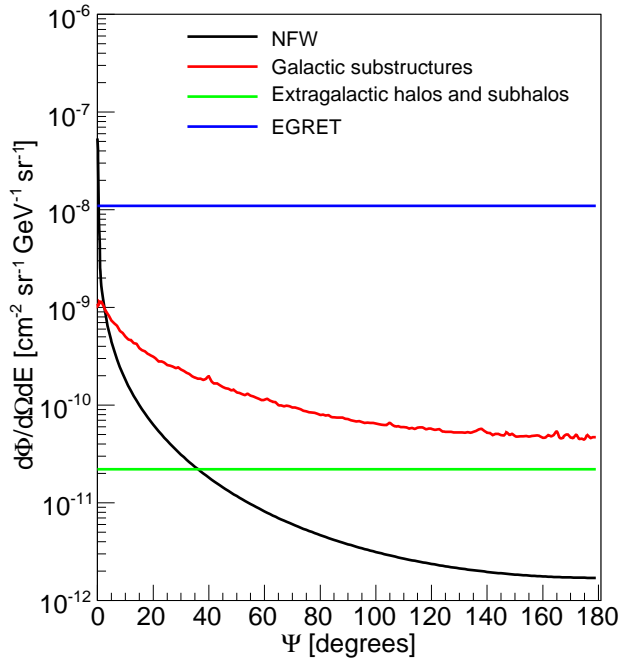}
\caption{\label{fig:Compare_flux_bz0ref}  Different contributions to the differential flux $ d\Phi/dEd\Omega $ [$ \mbox{cm}^{-2} \mbox{s}^{-1} \mbox{GeV}^{-1} \mbox{sr}^{-1} $] at 10 GeV as a function of the angle $ \Psi $ from the GC, in model  $B_{z_0,ref}$. The blue line corresponds to the {\small EGRET} estimate of the EGB as parametrized in Ref. \cite{Sreekumar:1997un}. The contribution from DM annihilation in the smooth host halo and its substructures are represented by the black and red curves, respectively. The latter is obtained by averaging over 10 different Monte Carlo realizations of the subhalos population. Bumps and wiggles are due to the contribution of the individual subhalos of the RES population. The green line represents the extra-galactic flux contributed by DM annihilation within extra-galactic halos and their substrutures. We have assumed $ m_\chi = 40 \mbox{ GeV} $, $ \sigma v = 3 \times 10^{-26} \mbox{cm}^3 \mbox{s}^{-1} $ and annihilations to $ b\bar{b}$.}
\end{figure}

\begin{figure}
\includegraphics[width=0.5\textwidth]{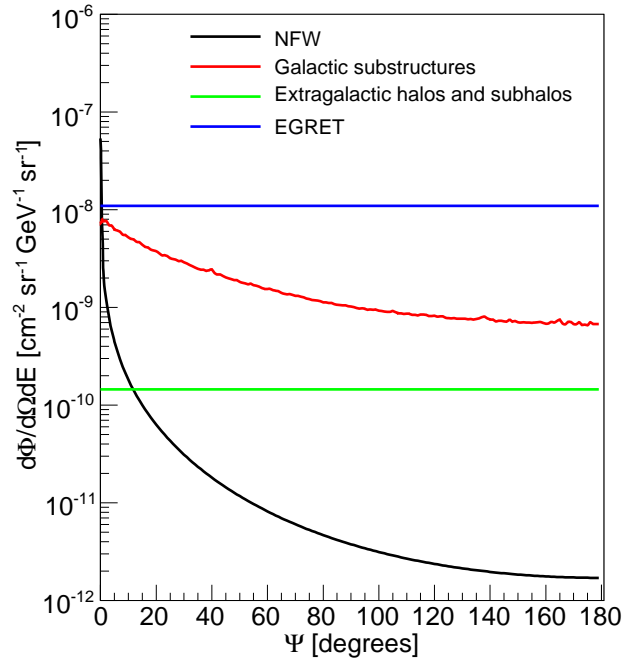}
\caption{\label{fig:Compare_flux_bzfref} 
Different contributions to the differential flux $ d\Phi/dEd\Omega $ at 10 GeV as a function of the angle $ \Psi $ from the GC in model $B_{z_f,ref}$. The curves represents the same quantities shown in Fig. \ref{fig:Compare_flux_bz0ref} but for $ B_{z_f,ref} $.}
\end{figure}

In Figures \ref{fig:Compare_flux_bz0ref} and \ref{fig:Compare_flux_bzfref}
we show the different contributions to the differential flux at 10 GeV
as a function of the angle $ \Psi $ from the GC in model  $B_{z_0,ref}$ and  
$B_{z_f,ref}$, respectively.

For $B_{z_0,ref}$ (Fig.~\ref{fig:Compare_flux_bz0ref}) the flux from
galactic substructures (red curve) dominates over the smooth NFW profile 
(black curve) at angular distances from the GC larger than about $ 3^\circ $ 
and it remains the dominant contribution at larger angles. The blue line 
labeled 'EGRET' represents the EGB inferred from  
{\small EGRET} measurements, equal to 
$ 1.10 \times 10^{-8} \mbox{cm}^{-2} \mbox{s}^{-1} \mbox{GeV}^{-1} \mbox{sr}^{-1} $ at 10 GeV, that can be used as a generous upper bound (the {\small Fermi LAT}
satellite will impose more stringent constraints, see next section).

Note that the NFW curve in Fig.~\ref{fig:Compare_flux_bz0ref} exceeds the 
{\small EGRET} EGB in the central degree, but not the flux of the bright 
gamma-ray source (J1746-2851) that has been identified by {\small EGRET} to 
be near - possibly coincident with - the GC, and it is thus consistent with 
observations.

For $ B_{z_f,ref} $, galactic substructures dominate the NFW signal at even 
smaller angular separations form the GC. In this case, the average flux for 
galactic substructures is larger than in the $B_{z_0,ref}$ case and almost 
matches the {\small EGRET} background.

\subsection{Extra-galactic contribution}
\label{sec:fourpointtwo}
The green line in Figures \ref{fig:Compare_flux_bz0ref} and 
\ref{fig:Compare_flux_bzfref} represents the average extra-galactic flux 
contributed by DM annihilation in extra-galactic halos, and their 
substructures, calculated with Eq.~\ref{eqn:aveflux}.
In model $ B_{z_0,ref} $ this flux turns out to be 
$ 2.20 \times 10^{-11} \mbox{cm}^{-2} \mbox{s}^{-1} \mbox{GeV}^{-1} \mbox{sr}^{-1} $, significantly smaller than in model $ B_{z_f,ref} $ (for which it is 
$ 1.45 \times 10^{-10} \mbox{cm}^{-2} \mbox{s}^{-1} \mbox{GeV}^{-1} \mbox{sr}^{-1} $) but still well below the galactic contribution and the {\small EGRET} 
constraint.
 
We stress the fact that in this work we have considered the {\small EGRET} 
background since it constitutes the best observational constraint available 
to date. This background is likely to be dominated by unresolved 
extra-galactic sources like blazars. Using the blazars Gamma-ray Luminosity 
Function derived from the blazars detected by {\small EGRET}, Ando et al. 
have shown that these sources should contribute only to 25-50\% of the total 
EGB \cite{Ando:2006cr}.
One therefore expects that, thanks to its superior sensitivity, 
{\small Fermi LAT} will be able to resolve a significant fraction of these 
sources \cite{Ando:2006mt}. As a result, the amplitude of the unresolved 
gamma-ray background will decrease, effectively lowering the blue line in Figs.
\ref{fig:Compare_flux_bz0ref}  and \ref{fig:Compare_flux_bzfref} in the
case of {\small Fermi LAT}.
This will have three advantages. First of all, a fainter gamma-ray background 
will increase the probability of detecting the gamma-ray annihilation signal 
produced by individual structures.
Second, the background from {\small Fermi LAT} will provide tighter 
constraints to our model for the DM annihilation, telling us if our 
description is too optimistic. Finally, with a larger number of resolved 
astrophysical sources, {\small Fermi LAT} will provide a better estimate of 
the blazar luminosity function.

The fact that the extra-galactic flux is contributed by both DM annihilation 
and blazars and because of the uncertainty on the latter contribution, one 
has the freedom of modifying the relative importance of blazars emission to 
the total EGB, keeping a good agreement with the data. Different choices for 
these relative contributions have been proposed in Ref. \cite{Ando:2006cr}.
They are listed in the first two columns of Tab. \ref{tab:percentages}. 
The same DM annihilation flux leads to a different value of 
$ f_{\mbox{\tiny{DM}}}^{\mbox{\tiny{EGRET}}} $ for {\small EGRET} and 
{\small Fermi LAT}, due to the lower {\small Fermi LAT} EGB.

It is particularly important to note that not all of the scenarios in 
Tab. \ref{tab:percentages} are physically plausible when implemented in our 
model.
For example, boosting the average galactic and extra-galactic DM annihilation
flux so that it will account for more than 61\% of the
{\small Fermi LAT} EGB, leads to a flux that exceeds the EGRET constraint
towards the GC (even excluding the central degree).
Thus, for $ B_{z_0,ref} $ we will only consider 
$f_{\mbox{\tiny{DM}}}^{\mbox{\tiny{Fermi}}}$ as large as 0.61, restricting
to the last two rows in Tab. \ref{tab:percentages}.
For the model $B_{z_f,ref}$, a similar argument rules out cases with 
$f_{\mbox{\tiny{DM}}}^{\mbox{\tiny{Fermi}}} > 0.80$.

\begin{table}
\begin{tabular}{c|c|c|c|c|c}
\hline
$ f_{\mbox{\tiny{blazars}}}^{\mbox{\tiny{EGRET}}} $ & $ f_{\mbox{\tiny{DM}}}^{\mbox{\tiny{EGRET}}}$ & $ f_{\mbox{\tiny{blazars}}}^{\mbox{\tiny{Fermi}}} $ & $ f_{\mbox{\tiny{DM}}}^{\mbox{\tiny{Fermi}}}$ & $ b.f. $ $ B_{z_0,ref} $ & $ b.f. $ $ B_{z_f,ref} $ \\
\hline
0.1 & 0.9 & 0.03 & 0.97 & ruled out & ruled out \\
0.3 & 0.7 & 0.20 & 0.80 & ruled out & 5.9 \\
0.5 & 0.5 & 0.39 & 0.61 & 49.8 & 4.2 \\ 
0.7 & 0.3 & 0.61 & 0.39 & 29.9 & 2.6 \\
\hline
\end{tabular}
\caption{\label{tab:percentages} $ f_{\mbox{\tiny{blazars}}}^{\mbox{\tiny{EGRET}}} $ and $ f_{\mbox{\tiny{DM}}}^{\mbox{\tiny{EGRET}}} $ indicate the contribution of gamma-ray from blazars and from DM annihilation respectively, to the EGB as measured by {\small EGRET}. $ f_{\mbox{\tiny{blazars}}}^{\mbox{\tiny{Fermi}}} $ and $ f_{\mbox{\tiny{DM}}}^{\mbox{\tiny{Fermi}}} $ represent the same quantities estimated for the {\small Fermi LAT} satellite in Ref. \cite{Ando:2006cr} using the blazar luminosity function derived from {\small EGRET} data. The last two rowes indicates the corresponding boost factors needed to bring the average annihation flux (galactic plus extra-galactic) to the relative percentage of the EGB value.}
\end{table}

\section{Angular power spectrum of  the gamma-ray unresolved signal}
\label{sec:five}

\subsection{Galactic contribution}

To compute the galactic gamma-ray angular power spectrum and investigate the 
relative importance of the different contributions, we separately analize the 
five maps of $ \Phi^{\rm cosmo} $ corresponding to the NFW, RES, UNRES, 
NFW$ \times $RES and NFW$ \times $UNRES contributions. In the maps, the value 
of $ \Phi^{\rm cosmo} $ is specified within angular bins of 
$ \Delta\Omega=9.57 \times 10^{-6} \mbox{sr} $.

Since we are interested in the contributions of DM halos and subhalos,
we mask out the region close to the GC and the Galactic plane in which the 
signal is dominated by gamma-rays produced by cosmic rays interacting with 
the interstellar galactic medium \cite{Strong:1998fr}. This background 
rapidly decreases with galactic latitude. For this reason we have used a  
composite mask consisting in a strip of $ 10^\circ $ above and below the 
Galactic plane and the squared area around the GC with coordinates 
$ |b| \le 30^\circ $ and $ |l| \le 30^\circ $. 
Outside the mask, the average galactic annihilation flux turns out to be
$ \langle d\Phi/dEd\Omega \rangle=8.79 \times 10^{-11} \mbox{cm}^{-2} \mbox{s}^{-1} \mbox{GeV}^{-1} \mbox{sr}^{-1} $ in the case of $ B_{z_0,ref} $ and 
$ \langle d\Phi/dEd\Omega \rangle=1.16 \times 10^{-9} \mbox{cm}^{-2} \mbox{s}^{-1} \mbox{GeV}^{-1} \mbox{sr}^{-1} $ for $ B_{z_f,ref} $, 
and it is dominated by the UNRES term.

To characterize the angular correlation signatures of the various 
contributions we compute the angular power spectrum of the different maps
with the {\bf HEALPix} 2.01 package 
\cite{Healpix,Gorski:1999rt}. Five {\ttfamily Healpix\_Map} objects are 
created covering the whole sky with 786432 pixels of constant area 
(corresponding to {\ttfamily N\_side} $ =2^8 $). Each pixel of the 
{\ttfamily Healpix\_Map} is filled with the corresponding flux 
$ d\Phi/dEd\Omega $ from the simulated maps.
The number of pixels is determined from the requirement that the area of the 
single bin is smaller than $ \Delta\Omega $.

The angular power spectrum, $C_\ell$, is computed from the 
spherical harmonic coefficients $ a_{\ell,m} $ of the gamma-ray flux angular 
fluctuation map as follows:
\begin{equation}
\left\langle \frac{d\Phi}{dEd\Omega} \right\rangle a_{\ell,m} =
\int d\Omega \left( \frac{d\Phi(\theta,\phi)}{dEd\Omega}-\left\langle 
\frac{d\Phi}{dEd\Omega} \right\rangle \right) Y_{\ell,m}(\theta,\phi)^{\ast}, 
\label{eqn:spectrum}
\end{equation}
\begin{equation}
C_\ell=\frac{\sum_{m=0}^{\ell} |a_{\ell,m}|^2}{2\ell+1},
\end{equation}
where $ Y_{\ell,m}(\theta,\phi) $ are the spherical harmonics and the 
integration in Eq. \ref{eqn:spectrum} extends to the unmasked area.
The angular spectra of the RES and NFW$ \times $RES maps have been obtained 
by averaging the spectra of the 10 independent maps corresponding to the
different Monte Carlo realizations of resolved subhalos.

We note in passing that the angular correlation signal does not 
depend on the particle physics scenario since it is evaluated with respect 
to the average flux (aside from a weak dependence on the minimum 
subhalo mass, which in turn depends on the properties of the DM particle). 
However, for the purpose of comparing the gamma-ray angular spectrum due to 
DM annihilation to that of extra-galactic astrophysical sources, a particular
particle physcis scenario has to be specified. In this case we refer to 
the reference values presented in Section \ref{sec:three}.

Our masking procedure induces spurious features in the NFW, UNRES and 
NFW$ \times $UNRES spectra that we smoothed out by averaging over 18 
multipoles.

The total angular spectrum is obtained by performing a sum of the 
aforementioned contributions and the cross-correlation terms: 
\begin{eqnarray}
\left\langle \frac{d\Phi^{\mbox{\tiny{tot}}}}{dEd\Omega} \right\rangle^2
C_\ell^{\mbox{\tiny{tot}}} & = &
\sum_i \left\langle \frac{d\Phi^{i}}{dEd\Omega} \right\rangle^2 
C_\ell^i + \\ \nonumber
& & \sum_{i,j} \left\langle \frac{d\Phi^{i}}{dEd\Omega} \right\rangle 
\left\langle \frac{d\Phi^{j}}{dEd\Omega} \right\rangle \sum_m a_{\ell,m}^i 
a_{\ell,m}^{\ast j},
\end{eqnarray}
with $ i,j \in $ [NFW, RES, UNRES, NFW$ \times $RES, NFW$ \times $UNRES].

\begin{figure}
\includegraphics[width=0.5\textwidth]{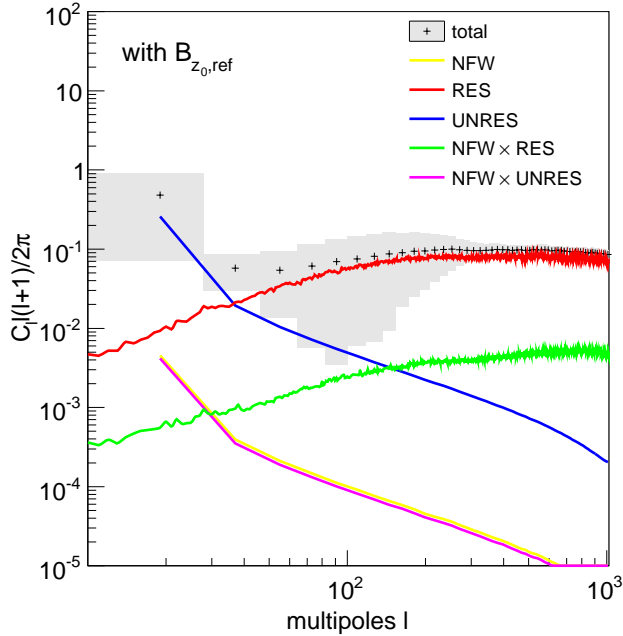}
\caption{\label{fig:CiElle_bz0ref} The curves represent the angular power spectrum $ C_\ell \ell(\ell+1)/2\pi $ of the different contributions to the galactic annihilation flux in the framework of model $ B_{z_0,ref} $. The labels identify the different contributions. Black crosses represent the angular spectrum of the total annihilation signal together with its  $ 1\sigma $ error (shaded areas).}
\end{figure}

\begin{figure}
\includegraphics[width=0.5\textwidth]{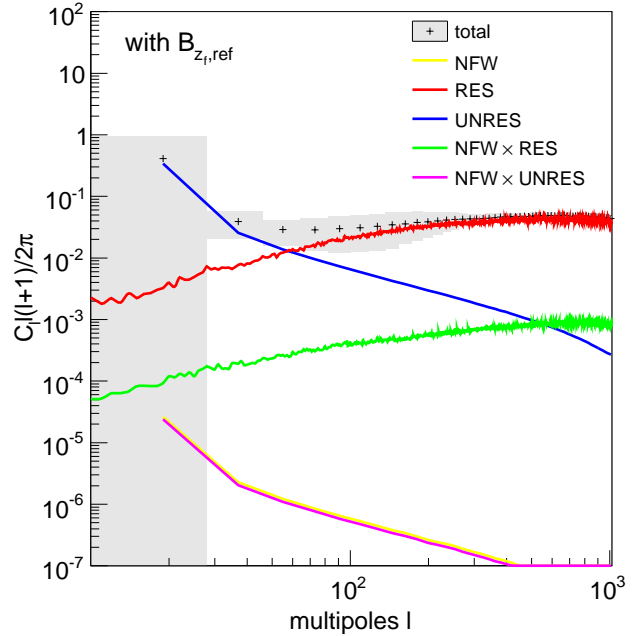}
\caption{\label{fig:CiElle_bzfref} 
The curves represent the angular power spectrum $ C_\ell \ell(\ell+1)/2\pi $ of the different contributions to the galactic annihilation flux in the framework of model $ B_{z_f,ref} $.}
\end{figure}

The results for model $ B_{z_0,ref} $ are plotted in 
Fig. \ref{fig:CiElle_bz0ref}, for which we also have taken into account the
pixel window function for the corresponding resolution of our maps 
\cite{Gorski:1999rt}.
The different curves in the plot represent the angular spectrum of 
each component weighted by the square of its relative contribution to the 
total annihilation flux $ \langle d\Phi^i/dEd\Omega \rangle^2 / 
\langle d\Phi^{\mbox{\tiny{tot}}}/dEd\Omega \rangle^2 $.
Cross-correlations are taken into account in the computation of the
total spectrum but are not shown in Figs. \ref{fig:CiElle_bz0ref} and
\ref{fig:CiElle_bzfref}.
Black crosses represent the spectrum of the total flux and the shaded areas
show the $1 \sigma$ error boxes. These are obtained by summing in quadrature 
the uncertainties related to the finite binsize and, if applicable, the 
scatter among the different Monte Carlo realizations of resolved subhalos.
The first contribution dominates at lower multipoles $ \ell $ while the
second becomes the main source of error at larger multipoles where the power
spectrum is mainly due to the RES contribution.
Poisson errors are not accounted for since they depend on the
observational setup and they will be discussed in the next Section for the
case of the {\small Fermi LAT} satellite.

The main contribution to the total flux is provided by the autocorrelation
of the resolved (red curve) and unresolved (blue curve) subhalos.
The former dominates at large multipoles while the latter dominates 
at $\ell<30$. The interplay between these two components is responsible for 
the minimum at $\ell \sim30$ in the total spectrum. All components involving 
the smooth DM distribution provide a negligible contribution to the total 
spectrum. 

Fig. \ref{fig:CiElle_bzfref} shows the angular spectrum of model 
$ B_{z_f,ref} $. The total spectrum and the different contributions are very 
similar to model $ B_{z_0,ref}$. The only difference is represented by 
the position of the minimum which is now around $ \ell \sim 50 $. This is
due to the contribution of the resolved subhalos (red curve) which is
smaller than in the previous case: subhalos in the  $ B_{z_f,ref} $ model 
are less concentrated than in  $ B_{z_0,ref} $ and consequently they contribute 
less to the spectrum at small angular scales.
We conclude that changing the concentration parameter significantly affects 
the mean annihilation flux but not the angular power spectrum of the signal.

Our model predictions depend on the assumption about the minimum subhalo mass, 
for which the theory does not provide very strong constraints.
In order to check the dependence of our results on the minimum subhalo mass, 
we have re-simulated the resolved subhalos excluding all structures below a 
given mass threshold. The spectrum does not change when excluding subhalos 
with a mass below $ 10^4 M_{\odot} $, showing that the angular correlation 
is mainly contributed by massive subhalos.

To check the reliability of our hybrid model, we have compared our 
results with those of a similar study recently carried out by Siegal-Gaskins
\cite{SiegalGaskins:2008ge} who used a pure Monte Carlo approach to compute 
the power spectrum of the annihilation signal produced by a population of 
subhalos in the MW above $ 10 M_{\odot} $ and  $ 10^7 M_{\odot}$, hence 
neglecting the contribution of smaller subhalos.
To compare our simulation approach to hers, we have modified our subhalo model 
to match that of Ref. \cite{SiegalGaskins:2008ge}. In particular we have 
re-simulated our galactic subhalo using a shallower mass function 
$dn/dM \propto M^{-1.9}$, assuming an Einasto density profile $ \rho_{halos} $
for subhalos, adopted a simplified power-law model for the concentration 
parameter with $ c(M) \propto M^{-0.138} $ and normalized the number of 
subhalos in the MW with a mass larger that $ 10^8 M_\odot$ to be 
$ N(M>10^8 M_\odot)=0.0064 (10^8 M_\odot/M_{\rm MW})^{-0.9} $ (this normalization
forces the number of MW subhalos to be five times smaller than with
the normalization assumed in Sec. \ref{sec:two} for the $ B_{z_0,ref} $ and 
$ B_{z_f,ref} $ cases).

The only residual difference between our model and that of Fig. 5 and 6 of
Ref. \cite{SiegalGaskins:2008ge} is the spatial subhalo distribution in the MW
which is more similar to the anti-biased model of 
Ref. \cite{SiegalGaskins:2008ge}.

\begin{figure}
\includegraphics[width=0.5\textwidth]{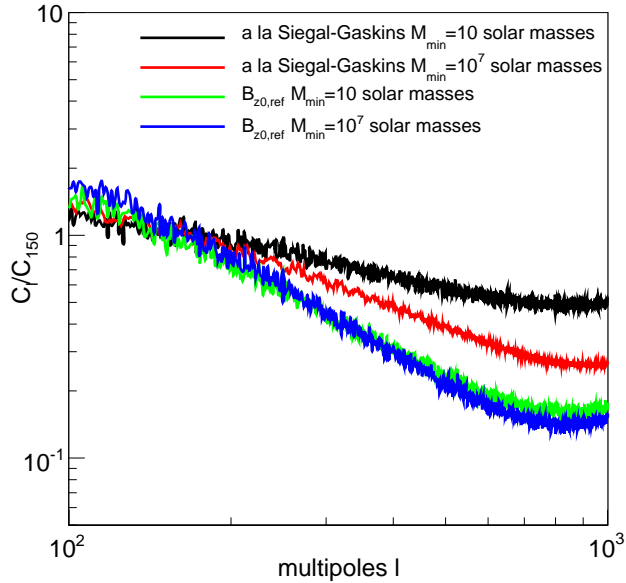}
\caption{\label{fig:CiElle_siegal_gaskins} Angular power spectrum $ C_\ell $, normalized to  $ C_{150} $ for the resolved subhalos modeled as in Ref. \cite{SiegalGaskins:2008ge}. The black (red) curve shows the case of a $ 10 M_\odot $ ($ 10^7 M_\odot $) mass cutoff. The green (blue) curve indicates the spectrum for the resolved subhalos in model $ B_{z_0,ref} $ for a mass cutoff of $ 10 M_\odot $ ($ 10^{7} M_\odot $).}
\end{figure}

The result of this exercise is shown in Fig. \ref{fig:CiElle_siegal_gaskins}. 
The curves shows the spectra normalized at $ C_\ell=150 $ for the various cases 
explored.
The spectra of the subhalos {\it a la} Siegal-Gaskins are consistent with 
Fig. 5 and 6 of Ref. \cite{SiegalGaskins:2008ge} for both the two mass 
thresholds considered ($ 10 \mbox{ } M_\odot $ (black curve) and $ 10^7 M_\odot$ 
(red curve)). The agreement is valid at the same time for the shape and the
amplitude of the signals.

The comparison with the spectrum of our subhalos in the case of 
$ B_{z_0,ref} $, shows that, for the same mass cut, our spectrum is steeper 
than in Ref. \cite{SiegalGaskins:2008ge}. 
This can be explained by noticing that small mass halos are more concentrated 
in Ref. \cite{SiegalGaskins:2008ge} than in our case, resulting in a larger 
power at small scales (i.e. at large multipoles).

Also the amplitude of the angular spectra predicted in models $ B_{z_0,ref} $ 
and $ B_{z_f,ref} $ results to be different than the same quantity in Fig. 5 of 
Ref. \cite{SiegalGaskins:2008ge}, even when the different subhalo spatial 
distribution is considered: our spectra are characterized by a lower
amplitude, the result of having considered subhalos with masses lower than 
10 $ M_\odot $. They contribute to the total annihilation flux but their 
distribution is almost homogeneous around the observer, hence decreasing the 
spectral amplitude.

Finally, as we have already pointed out, the results presented here are robust 
to the choice of the maximal simulation distance. We have checked that 
resolving halos at distances larger than $ 3 \times d_{max}(M) $ has very 
little impact on the angular power spectrum.

\subsection{Extra-galactic contribution}
\label{subsec:extragalactic}

Our analysis for the angular spectrum of the extra-galactic annihilation flux
is similar to the one originally developed by Ando et al. in 
Ref. \cite{Ando:2005xg} and extended in \cite{Ando:2006cr} to account for 
substructures. 
Like in our model, these authors assume that DM halos and subhalos
have NFW density profiles and that subhalos trace the smooth underlying mass 
distribution of the host halo.
Moreover they assume that the mean number of subhalos within a host halo
of mass $M$, $\langle N|M \rangle$, scales as $ \propto M^{\alpha} $ and 
they consider two extreme cases, $ \alpha=1 $ and $ \alpha=0.7 $
\cite{Ando:2006cr}. In this work we only consider the first case
to match the subhalo mass function of our model. 

We notice that Ando et al. \cite{Ando:2006cr} do not account for the 
contribution of subhalos by integrating over their mass function, as in the 
second term of Eq. \ref{eqn:substructures}.
Instead, they consider a gamma-ray subhalo spectrum averaged over its mass 
function and multiplied by the mean comoving density of subhalos.
These model details hardly affect the angular power spectrum since one 
considers fluctuations about the mean flux.
While we will quantify the differences between the two approaches in a 
future work in which the angular power spectrum and the mean flux will be 
computed self-consistently, here, for the purpose of discussing the angular 
power spectrum, we adopt the same approach as in Ref. \cite{Ando:2006cr} and 
we will use their very same angular power spectra shown in Fig. 5 of their 
paper for both the DM annihilation and the blazars contribution to the 
extra-galactic flux.

We do not plot the contribution of the DM and blazars to the angular spectrum 
of the extra-galactic gamma-ray background in Figs. \ref{fig:CiElle_bz0ref}  
and \ref{fig:CiElle_bzfref} to avoid confusion. 
Instead, in the next Section we show how well these spectra will be measured
by {\small Fermi LAT} and compare them to the galactic contributions.

\section{{\small Fermi LAT} gamma-ray angular spectra}
\label{sec:six}

In this section we apply our model to predict the angular power spectrum that
{\small Fermi LAT} is expected to observe and discuss the possibility of 
indirect DM detection through characteristic signatures in the spectrum.

\begin{figure*}
\includegraphics[width=0.45\textwidth]{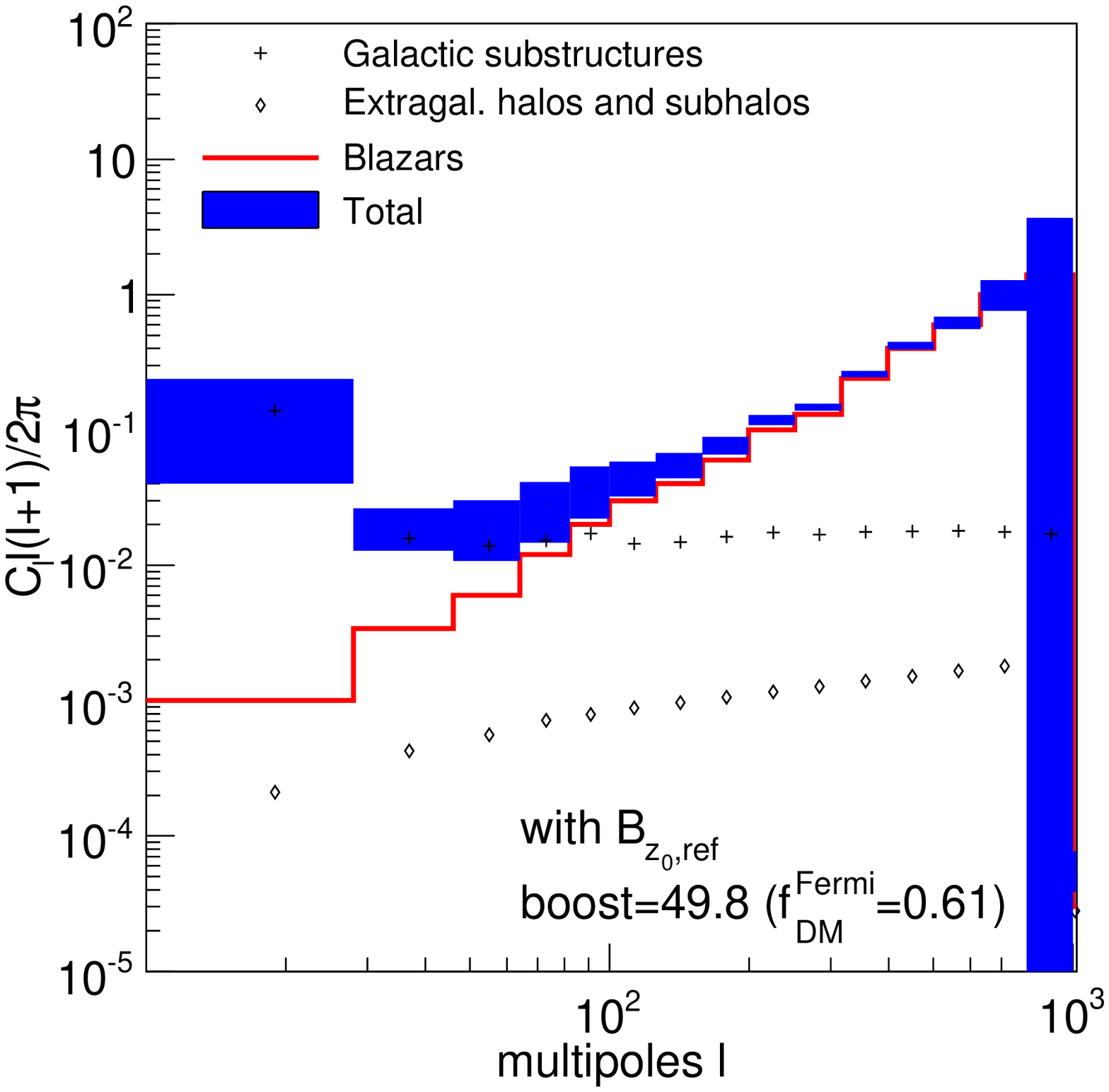}
\includegraphics[width=0.45\textwidth]{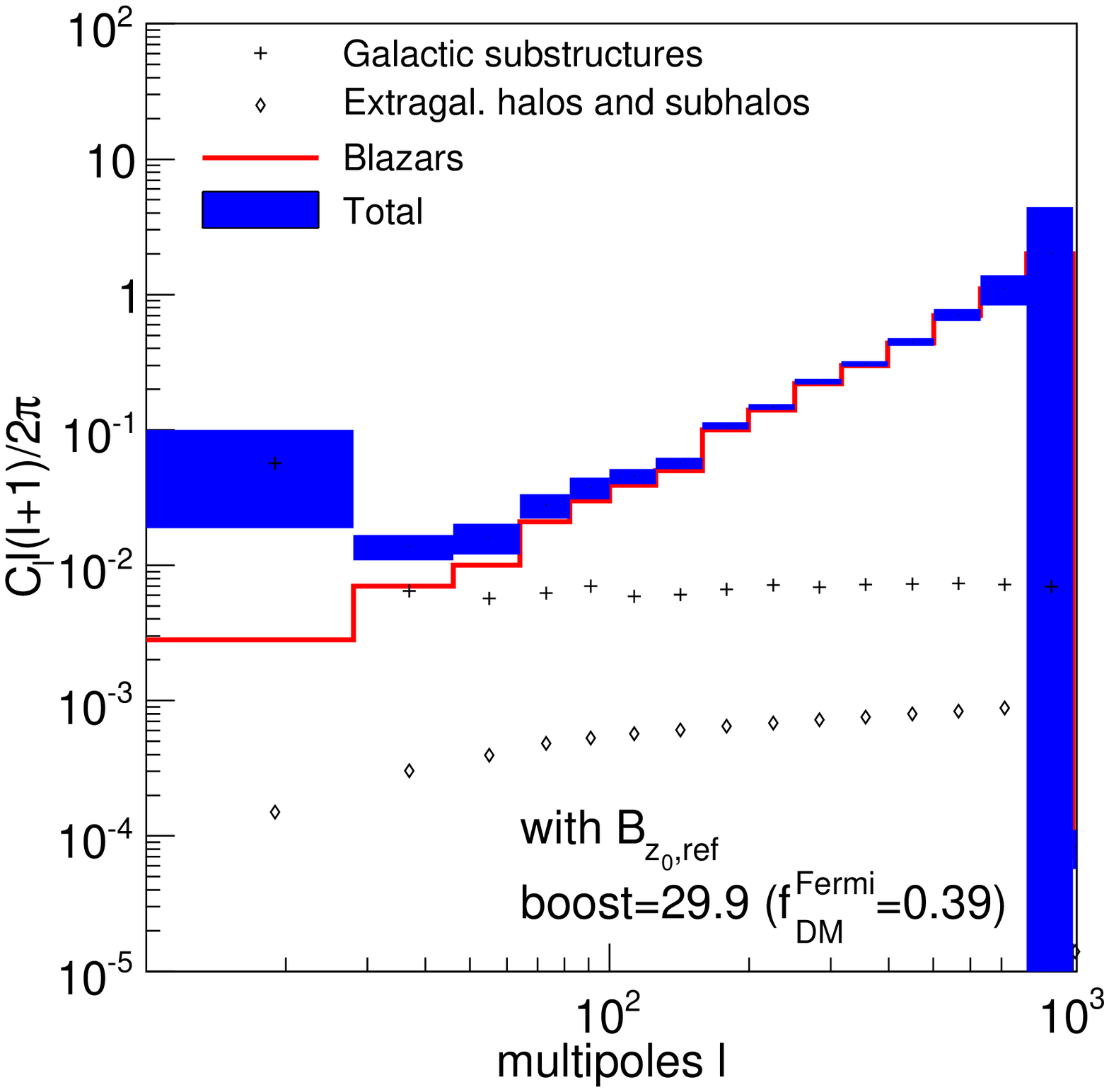}
\caption{\label{fig:CiElle_bz0ref_percentage} Angular power spectrum of gamma-ray anisotropies in the case that DM annihilations contribute to 61\% (left) and 39\%(right) of the total average flux of the EGB as estimated for the {\small Fermi LAT} satellite. The plus signs $ + $ indicate the galactic component following the $ B_{z_0,ref} $ model (see Fig. \ref{fig:CiElle_bz0ref}). The diamonds $ \diamond $ refer to DM annihilating in extra-galactic, unresolved DM halos and subhalos and the spectrum is taken from Ref. \cite{Ando:2006cr}. The red line represents the angular spectrum of blazars as described in Ref. \cite{Ando:2006cr}. All these three components are normalized to their average contribution to the total flux $ \langle d\Phi^i/dEd\Omega \rangle^2/ \langle d\Phi^{\rm tot}/dEd\Omega \rangle ^2 $. The $ 1 \sigma $ error of the total spectrum is indicated by the blue boxes and it is obtained propagating the errors from the galactic and the extragalactic DM components, assuming no errors for the blazars.}
\end{figure*}

The contribution of the different components to the total angular spectrum
of the gamma-ray flux is shown in Fig. \ref{fig:CiElle_bz0ref_percentage} 
for the $ B_{z_0,ref} $ case. The red histogram indicates the angular spectrum
of anisotropies in the blazar emissivity (see Fig. 5 of 
Ref. \cite{Ando:2006cr}). 
Black crosses represents the contribution of galactic substructures that, 
as we have shown, dominate over the smooth NFW signal.
Open diamonds indicate the contribution of extra-galactic halos and their 
substructures. All the different contributions are normalized to the square
of the ratio between the average flux of that particular contribution to the
average total flux $ \langle d\Phi^i/dEd\Omega \rangle^2 / 
\langle d\Phi^{\mbox{\tiny{tot}}}/dEd\Omega \rangle^2 $.

Poissonian errors in each bin of the sky maps are due to the small statistics,
i.e. the limited number of gamma-ray photons from annihilation collected 
during observations. The $ 1\sigma $ error is computed, following 
Ref. \cite{Ando:2006cr}, as follows:
\begin{equation}
\delta C_l=\sqrt{\frac{2}{(2l+1) \Delta l f_{\mbox{\tiny{f.o.v.}}}}}
\left( C_l + C_l^b + \frac{C_N}{W_l^2} \right),
\label{eqn:poissonian_error}
\end{equation}
where $ \Delta l=18 $ and $ 4\pi f_{\mbox{\tiny{f.o.v.}}}=9.706 \mbox{ sr} $
is the area outside the mask. $ C_N $ is the power spectrum 
of the photon noise $ C_N=4\pi f_{\mbox{\tiny{f.o.v.}}}/N_{\rm EGB} $
that depends on the instrument characteristics and integration time 
($ N_{\rm EGB} $ is the number of photons of the EGB) and is independent from 
$ \ell $. We have assumed an effective area constant with energy of 
$ 10^4 \mbox{cm}^2 $ and an exposure time of 1 year. $ W_\ell$ is the window 
function of a Gaussian point spread function 
$ W_\ell=\exp(-\ell^2 \sigma_b^2/2) $ 
that we compute for the {\small Fermi LAT} angular resolution 
$ \sigma_b=0.115^\circ $.
The angular spectrum of the background $ C_\ell^b $ is not uniquely defined but
depends on the signal one wants to detect: for the extra-galactic DM 
component, the background is represented by blazars and by the galactic 
annihilations. While, viceversa, for the galactic DM component, blazars
and extra-galactic annihilations play the role of background.

\begin{figure*}
\includegraphics[width=0.32\textwidth]{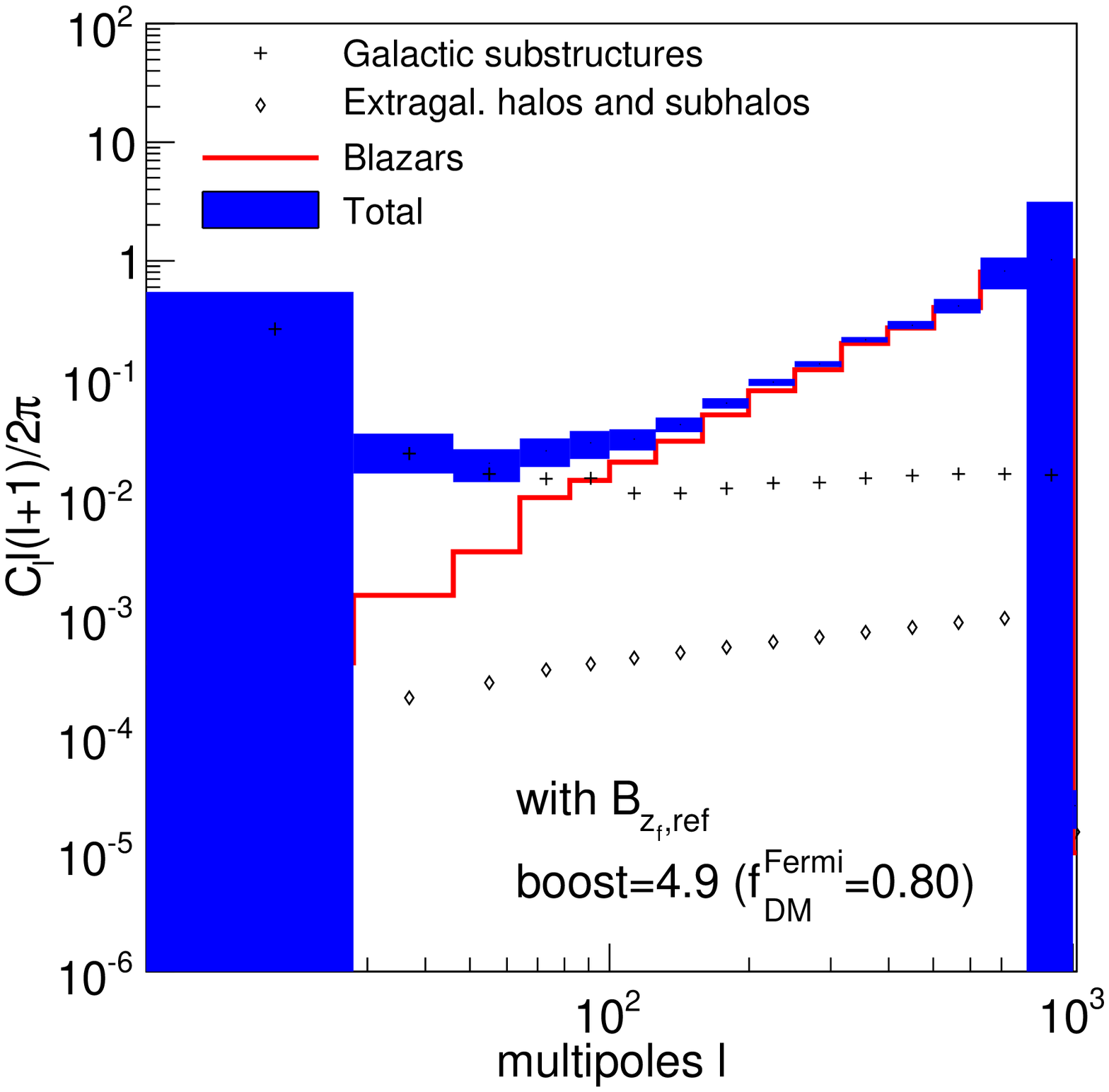}
\includegraphics[width=0.32\textwidth]{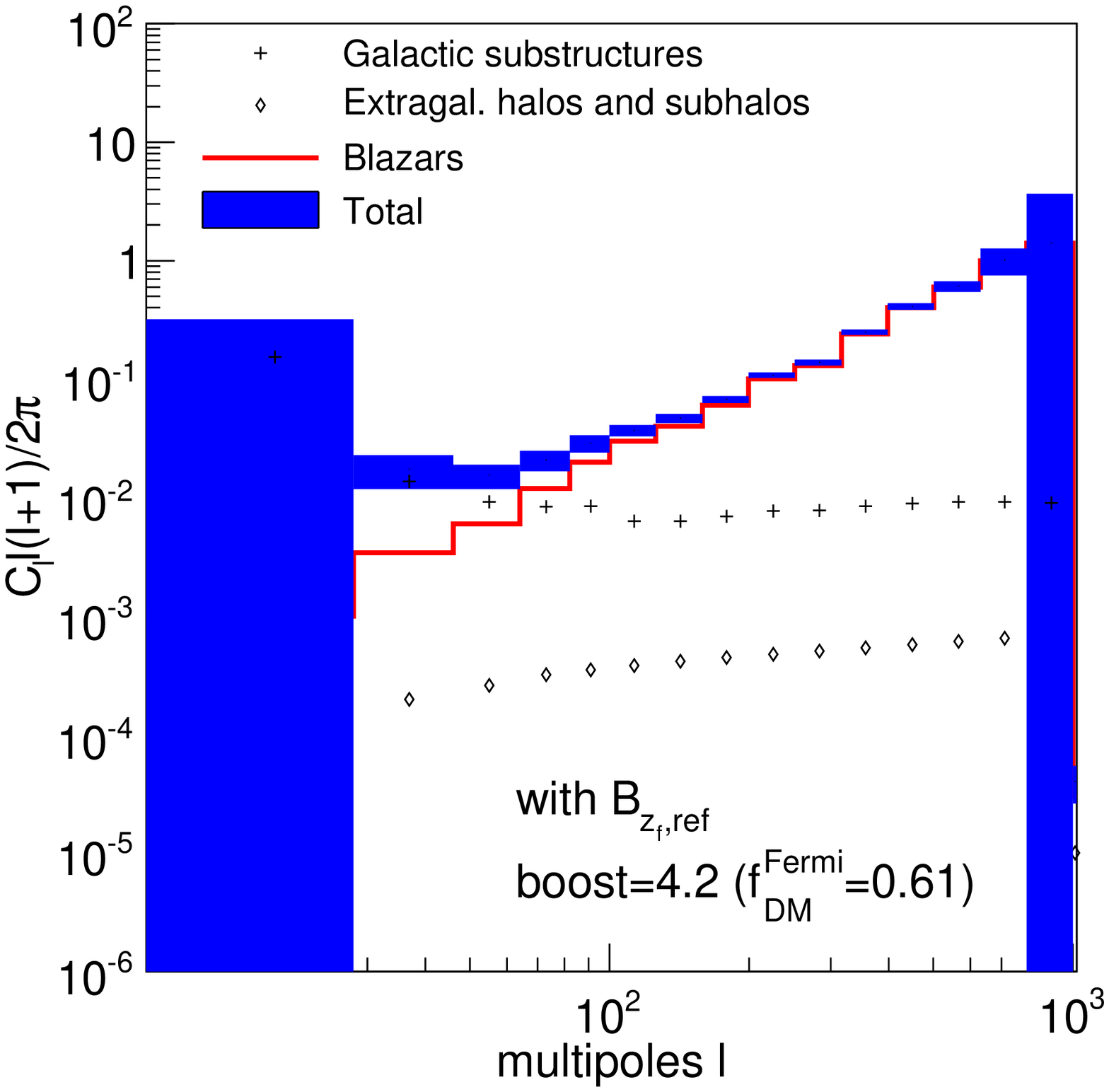}
\includegraphics[width=0.32\textwidth]{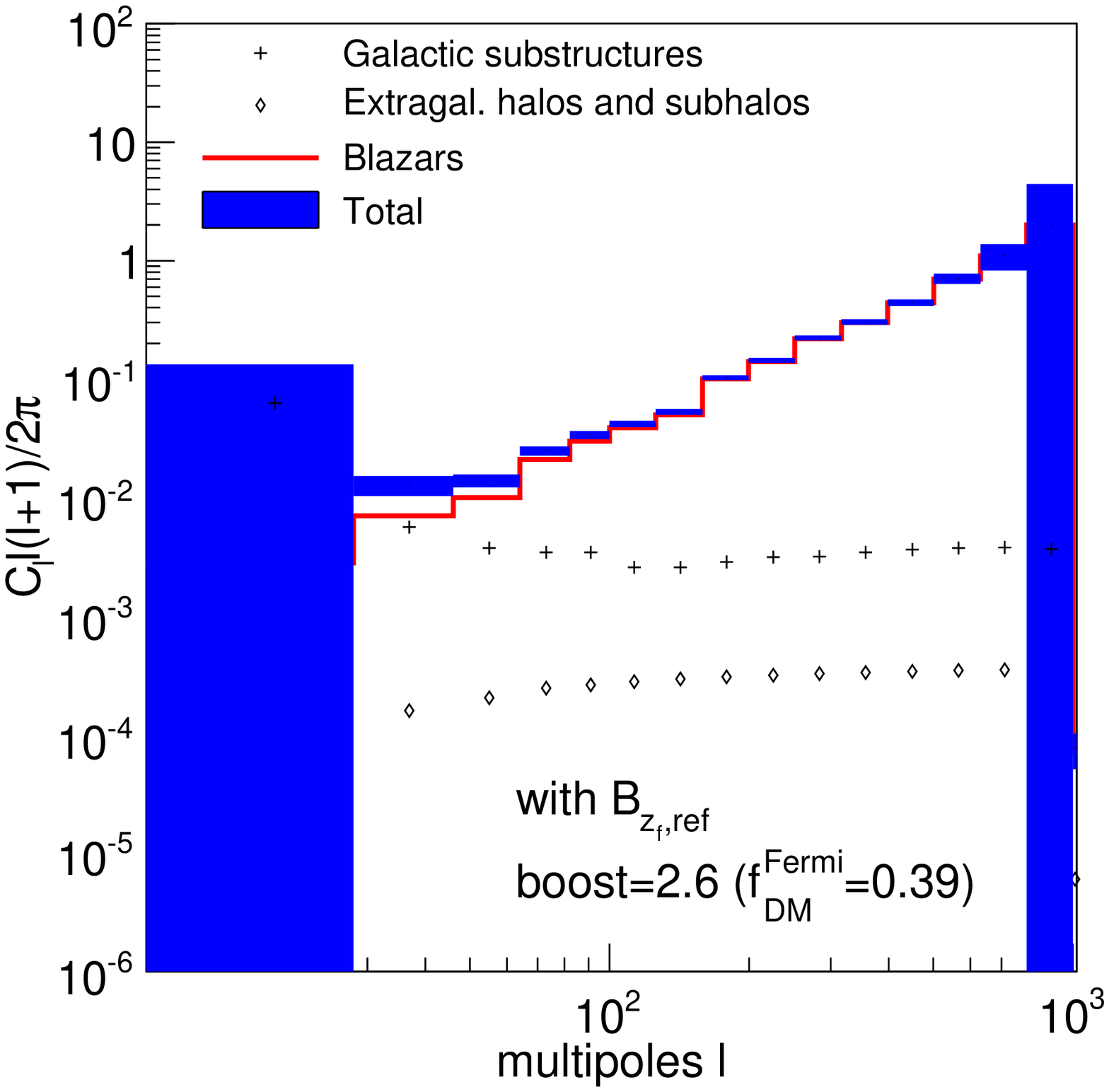}
\caption{\label{fig:CiElle_bzfref_percentage} Angular power spectrum of gamma-ray anisotropies in the case that DM annihilations contribute to 80\% (left), 61\%(center) and 39\% (right) of the total average flux of the EGB as estimated for the {\small Fermi LAT} satellite. The curves and the error boxes have the same meaning of Fig. \ref{fig:CiElle_bz0ref_percentage} but refer here to the $ B_{z_f,ref} $ case.}
\end{figure*}

The total signal (shown as crosses in Fig. \ref{fig:CiElle_bz0ref_percentage} 
and Fig. \ref{fig:CiElle_bzfref_percentage}) is obtained by adding all 
components weighted by their relative contribution to the total gamma-ray 
flux $ \langle d\Phi^i/dEd\Omega \rangle^2/ \langle d\Phi^{\rm tot}/dEd\Omega \rangle ^2 $ including the cross-correlation term involving extra-galactic
halos and blazars. We have ignored the cross-correlation between the galactic 
subhalos and the blazars since they have independent spatial distributions. 
The total errors (blue boxes) account for the Poisson noise of both the 
galactic and extra-galactic DM component, to which we summed the binsize and 
scatter among Monte Carlo realizations. The Poissonian noise becomes the
main source of errors only al large multipoles.
Finally, we have assumed that the angular power spectrum of blazars is 
known without errors.

In the particle physics scenario considered in this paper and for the 
$ B_{z_0,ref} $ model, the average flux $ \langle d\Phi/dEd\Omega \rangle $ 
produced by DM annihilations within and outside our galaxy is less than 1\% of 
the EGB estimated for {\small Fermi LAT}. We can therefore boost up this 
contribution by increasing the cross section of the DM candidate. 
This has the effect of changing the values of 
$f_{\mbox{\tiny{blazars}}}^{\mbox{\tiny{Fermi}}} $ and 
$ f_{\mbox{\tiny{DM}}}^{\mbox{\tiny{Fermi}}}$, i.e.
the relative contribution of blazars and DM to the total EGB.
The two panels in Figure \ref{fig:CiElle_bz0ref_percentage}
refer to  $ f_{\mbox{\tiny{DM}}}^{\mbox{\tiny{Fermi}}} = 0.61$
(left) and  $ f_{\mbox{\tiny{DM}}}^{\mbox{\tiny{Fermi}}} = 0.39$
(right). The corresponding boosting factors are shown in the plots.
Larger boosting factors (corresponding to larger  
$ f_{\mbox{\tiny{DM}}}^{\mbox{\tiny{Fermi}}} $) are excluded since they would 
correspond to models already excluded by the current {\small EGRET} constraint.

In both plots the contribution of the galactic signal to the angular power 
spectrum largely dominates the extra-galactic component at all multipoles. 
As a consequence, the two main contributions to the angular spectrum are 
provided by the blazars and the galactic annihilation signal. The former 
dominates at large multipoles whereas the latter dominates at low $\ell$. 
The position of the crossover depends on the boosting factor, but even in the 
less favorable case of  $ f_{\mbox{\tiny{DM}}}^{\mbox{\tiny{Fermi}}} = 0.39$
the DM annihilation signature can be clearly seen as a turnover in the power 
spectrum at $\ell < 30$.

These considerations remain valid for the $ B_{z_f,ref} $ model whose angular 
power spectrum is shown in Fig. \ref{fig:CiElle_bzfref_percentage}. In this 
case, however, the crossover is found at smaller multipoles, making more 
difficult to detect the contribution for DM annihilation when  
$ f_{\mbox{\tiny{DM}}}^{\mbox{\tiny{Fermi}}} = 0.39$ because of the large 
errorbars. On the other hand, with the $ B_{z_f,ref} $ model one can increase 
the boost factor up to $ f_{\mbox{\tiny{DM}}}^{\mbox{\tiny{Fermi}}}=0.80 $ 
without ending up with an unphysical scenario already excluded by 
{\small EGRET}.

\section{Discussion and conclusions}
\label{sec:seven}
In this paper we have modeled for the first time the angular power spectrum 
of the diffuse gamma-ray signal at high galactic latitude taking into account 
both the DM annihilation in substructures within our Galaxy and annihilation 
within extra-galactic halos and their subhalos, also including the 
contribution from extra-galactic objects like blazars.
The purpose is to understand whether the gamma-ray satellite 
{\small Fermi LAT} will be able to unambiguously detect the signature of DM 
annihilation that, in the $\Lambda$CDM scenario, preferentially takes place 
in the central regions of DM halos and in their hierarchically-nested 
substructures. 

For this purpose we have developed an hybrid approach in which the 
annihilation signal is computed using two different techniques.
First, we perform a numerical integration to compute the contributions 
to the gamma-ray flux from the smooth galactic halo, from the unresolved 
galactic substructures and, finally, from extra-galactic DM halos and their 
substructures. Second, we use Monte Carlo techniques to account for
the nearest galactic subhalos, that show up in the mock gamma-ray maps
as individual gamma-ray sources.

We have explored two different subhalo
models corresponding to different prescriptions for the mass concentration 
within the halo, $ B_{z_0,ref} $ and $ B_{z_f,ref}$.
 
The EGB is computed by adding the annihilation signal from extra-galactic 
halos and that from high energy astrophysical sources like blazars. The 
relative contributions of the two signals is treated as a free parameter. 
The only constraint is that the total gamma-ray signal, galactic annihilation 
foreground and EGB, does not exceed the current limit of {\small EGRET}.

Our main conclusions are:

\begin{itemize}

\item The annihilation of the smooth galactic halo dominates over 
that of galactic substructures only within a few degrees from the GC. The 
exact value depends on the subhalos concentration and ranges from 
$< 1^{\circ}$ in model $ B_{z_f,ref} $,  to $\sim 3^{\circ}$ in model
$ B_{z_0,ref} $.
Moreover, the gamma-ray flux from extra-galactic halos is fainter than the 
galactic signal at all angles. This is true both for $ B_{z_0,ref} $ and
$ B_{z_f,ref} $.

\item To predict the annihilation flux we have assumed a favourable
particle physics setup, with a mass of 40 GeV, a cross section 
$\sigma v=3\times 10^{-26} \ {\rm cm}^{-3} {\rm s}^{-1}$, and annihilations 
to $b \overline{b}$. With this choice, model $ B_{z_0,ref} $ 
accounts for only 0.8\% of the total unresolved EGB that will be measured 
by {\small Fermi LAT}. Larger fluxes are possible for smaller masses and
larger cross sections, provided that the total flux does not exceed the 
current measurement of the EGB by {\small EGRET} and still allowing blazars 
to contribute to a significant fraction of the flux. Similar considerations 
also apply to model $ B_{z_f,ref} $. The largest boost factor compatible with 
the data is of order 100, and it could be achieved e.g. through the so-called 
Sommerfeld enhancement or in some rather fine-tuned supersymmetric scenarios
~\cite{Profumo:2005xd,Sommerfeld}.

\item  The angular power spectrum predicted by our model is reliable and 
robust. To test its reliability we compared our results with those recently 
published in Ref. \cite{SiegalGaskins:2008ge}, who performed a similar 
analysis considering, however, only galactic substructures and using a full 
Monte Carlo approach. When the same halo model is considered, we find fair 
agreement with Ref. \cite{SiegalGaskins:2008ge}.
To test the robustness of our method we have verified that our results
do not change significantly when we increase the volume within which  
we simulate the distribution of galactic subhalos beyond a well defined 
reference value.

\item The angular power spectrum of the galactic gamma-ray photons produced 
by DM annihilation is dominated by subhalos. Contributions involving the 
DM particles in the smooth galactic halo are negligible at all multipoles. 
At low multipoles (large angles) the spectrum is dominated by 
unresolved halos whose annihilation flux appear in the gamma-ray sky map
as a smooth foreground (see e.g. Figs. 6 and 7 of Ref. \cite{Pieri:2007ir}). 
On the other hand, nearby resolved clumps dominate the power at small angular 
separations (large multipoles). These results are robust in the sense that 
they do not change appreciably when adopting different prescriptions for the 
concentration parameters.
This is not surprising since the power spectrum, which refers to fluctuations
about the mean flux, should not depend appreciably on the subhalo structure.

\item  Subhalos of masses below $10^{4} M_{\odot}$ give a small contribution 
to the angular spectrum of the galactic annihilation flux.

\item The signature of DM annihilations in the angular power spectrum
can be found only for an optimistic particle physics setup.
In fact, for our benchmark particle physics scenario (without any boost 
factor) the power spectrum of the unresolved gamma-ray background would
be completely dominated by blazars. However, boosting up the particle 
physics factor without exceeding the background that {\small EGRET} has
estimated, we find that the power spectrum is dominated by the DM component 
at low multipoles while blazars determine the spectrum at small angular 
scales. The turn-over depends on the prescription for the concentration of 
subhalos, occuring at smaller scales in the case of $ B_{z_0,ref} $.

\item The angular power spectrum of DM annihilations is largely dominated,
at all scales, by the galactic signal. It depends mostly on the relative
importance of the average galactic flux compared to the extra-galactic
DM signal which, in our models, favours the first term, both for 
$ B_{z_0,ref} $ and $ B_{z_f,ref} $.

\item We find that a 1-year all-sky survey with {\small Fermi LAT}, now 
operational, may be able to spot the annihilation signature of DM in the 
angular power spectrum of the unresolved gamma-ray flux. This signature is 
provided by the up-turn in the power spectrum which, at low multipoles, is 
dominated by the DM annihilation signal.
Our results partially contradict those of Ref. \cite{Ando:2006cr} who 
did not account for the galactic contribution to the angular power spectrum
which, instead, completely obliterate the one provided by annihilations in
extra-galactic halos.

\item Our results seem to confirm those of Ref. \cite{SiegalGaskins:2008ge}
since the angular spectrum measured by {\small Fermi LAT} should indeed 
provide an indirect detection from DM annihilation within our Galaxy, although 
it is doubtful that such measurement will be able to constrain the 
properties of the subhalo population. 
However, we should point out that the subhalo description adopted in 
Ref. \cite{SiegalGaskins:2008ge} would boost up the extra-galactic DM 
annihilation flux to a value of  
$ 3.60 \times 10^{-10} \mbox{cm}^{-2} \mbox{s}^{-1} \mbox{GeV}^{-1} $ 
comparable to, if not larger than, the galactic annihilation flux. On one side 
this fact increases the possibility of detecting the DM contribution to the
angular spectrum. On the other hand it will be hard to disentangle 
galactic and extra-galactic contributions.
\end{itemize}

As a remark we stress that the results of the recent {\it Via Lactea II} 
\cite{Diemand:2008in} and the 
{\it Aquarius} \cite{Springel:2008by,Springel:2008cc} numerical experiments, 
both with improved resolution, have now updated our knowledge on the subhalo 
distribution function and concentration parameters.
The differences in the subhalos extracted from these simulations, whose main
characteristics are listed in Table \ref{tab:smodels}, may change the 
angular spectrum of the DM annihilation flux although we have shown 
that changing subhalo model significantly affects the gamma-ray flux but has 
less impact on the spectral analysis.
We will investigate this issue thoroughly in a forthcoming paper. Here we 
just report the result of a preliminary analysis in which we have computed 
the angular power spectrum of the galactic annihilation signal obtained when 
subhalos are modeled according to {\it Via Lactea II}  and {\it Aquarius}.
In both cases we find that the angular spectrum is flatter, especially at 
small $\ell$. This is not surprising since, in both experiments, the mass 
distribution is less clumpy and the subhalo radial distribution is flatter 
than in the model adopted in the present paper.
A flatter spectrum, combined with a fainter annihilation flux due to the 
reduced number of substructures, would reduce the up-turn in the angular 
spectrum, making it more challenging to detect the DM annihilation feature. 
The significance of this effect depends on the boosting factors that, 
because of the dimmer flux, could be further increased without exceeding the 
measured gamma-ray background.

Finally, being tuned to the results of N-body simulations, our description of 
the galactic and extra-galactic subhalo population does not account for the
effect of baryons in the evolution of DM structures. As a consequence our
computation of the angular spectrum does not account for the effect, e.g.,
of Intermediate Mass Black Holes, which may be present in our Galaxy and
which correlation properties have been recently studied in 
Ref. \cite{Taoso:2008qz}.

\section*{Acknowledgments}
\noindent 
We would like to thank S. Ando, M. Gustafsson, E. Komatsu and J. Siegal-Gaskins 
for very useful comments on the manuscript. We would also like to thank 
M. Taoso for useful discussions and S. Ricciardi and E. Hivon for the help 
with the {\bf HEALPix} package. LP would like to thank the European Network of 
Theoretical Astroparticle Physics ENTApP ILIAS/N6 under contract number 
RII3-CT-2004-506222 for financial support. EB thanks the Institut 
d'Astrophysique de Paris for the kind hospitality and acknowledge financial 
contribution from contracts ASI-INAF/TH-014 and ASI-INAF/TH-027.

\end{document}